\documentclass[aps,prl,twocolumn,showpacs,amsmath,amssymb]{revtex4-1}
\usepackage{amsmath}
\usepackage{graphicx}
\usepackage{adjustbox}
\usepackage{subfigure}
\usepackage{epstopdf}
\usepackage{color}
\usepackage{multirow}
\usepackage{setspace}
\usepackage{overpic}
\usepackage{amssymb}
\usepackage{placeins}

\usepackage[colorlinks,urlcolor=blue, citecolor=blue,linkcolor=blue] {hyperref}
\usepackage{lineno}
\usepackage{bm}
\usepackage{rotating}
\usepackage[utf8]{inputenc}
\let\oldequation\equation
\let\oldendequation\endequation
\renewenvironment{equation}
 {\linenomathNonumbers\oldequation}
 {\oldendequation\endlinenomath}

\newcommand{\BESIIIorcid}[1]{\href{https://orcid.org/#1}{\hspace*{0.1em}\raisebox{-0.45ex}{\includegraphics[width=1em]{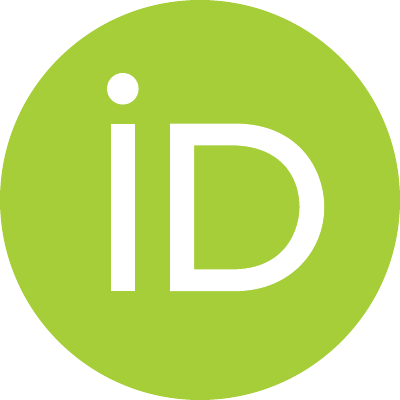}}}}

\begin{document}
	\renewcommand{\thesection}{\Roman{section}}

\title{\bf \boldmath
Measurements of the absolute  branching fractions of $D^0\to\gamma \bar K^{*0}$ and $D^0\to\gamma \phi$
}

\author{M.~Ablikim$^{1}$\BESIIIorcid{0000-0002-3935-619X},
	M.~N.~Achasov$^{4,c}$\BESIIIorcid{0000-0002-9400-8622},
	P.~Adlarson$^{83}$\BESIIIorcid{0000-0001-6280-3851},
	X.~C.~Ai$^{89}$\BESIIIorcid{0000-0003-3856-2415},
	C.~S.~Akondi$^{31A,31B}$\BESIIIorcid{0000-0001-6303-5217},
	R.~Aliberti$^{39}$\BESIIIorcid{0000-0003-3500-4012},
	A.~Amoroso$^{82A,82C}$\BESIIIorcid{0000-0002-3095-8610},
	Q.~An$^{79,65,\dagger}$,
	Y.~H.~An$^{89}$\BESIIIorcid{0009-0008-3419-0849},
	Y.~Bai$^{63}$\BESIIIorcid{0000-0001-6593-5665}
% ...
	\\
	\vspace{0.2cm}
	(BESIII Collaboration)\\
	\vspace{0.2cm} {\it
		$^{1}$ Institute of High Energy Physics, Beijing 100049, People's Republic of China\\
% ...
$^{4}$ Budker Institute of Nuclear Physics SB RAS (BINP), Novosibirsk 630090, Russia\\
% ...
$^{31}$ INFN Sezione di Ferrara, (A)INFN Sezione di Ferrara, I-44122, Ferrara, Italy; (B)University of Ferrara, I-44122, Ferrara, Italy\\
% ...
$^{39}$ Johannes Gutenberg University of Mainz, Johann-Joachim-Becher-Weg 45, D-55099 Mainz, Germany\\
% ...
$^{63}$ Southeast University, Nanjing 211100, People's Republic of China\\
% ...
$^{65}$ State Key Laboratory of Particle Detection and Electronics, Beijing 100049, Hefei 230026, People's Republic of China\\
% ...
$^{79}$ University of Science and Technology of China, Hefei 230026, People's Republic of China\\
% ...
$^{82}$ University of Turin and INFN, (A)University of Turin, I-10125, Turin, Italy; (B)University of Eastern Piedmont, I-15121, Alessandria, Italy; (C)INFN, I-10125, Turin, Italy\\
		$^{83}$ Uppsala University, Box 516, SE-75120 Uppsala, Sweden\\
% ...
$^{89}$ Zhengzhou University, Zhengzhou 450001, People's Republic of China\\
% ...
$^{\dagger}$ Deceased\\
% ...
$^{c}$ Also at the Novosibirsk State University, Novosibirsk, 630090, Russia\\
% ...
}}

\begin{abstract}
By analyzing a sample of 20.3 fb$^{-1}$ of $e^+e^-$ annihilation data  collected at the center-of-mass energy of 3.773~GeV with the BESIII detector, we have made a first measurement of the absolute branching fractions of the radiative decays
$D^0\to\gamma \bar K^{*0}$ and $D^0\to\gamma\phi$ to be $(3.81 \pm 0.18_{\rm stat.} \pm 0.20_{\rm syst.})\times 10^{-4}$ and $(2.51 \pm 0.44_{\rm stat.} \pm 0.11_{\rm syst.})\times 10^{-5}$, respectively.
 The statistical significances of $D^0\to\gamma \bar K^{*0}$ and
$D^0\to\gamma \phi$ are $26.8\sigma$ and $7.9\sigma$, respectively.
The obtained branching fractions are consistent with the corresponding world average values. In addition, the measured $C\!P$
asymmetry $\mathcal{A}_{C\!P}(D^0\to\gamma \bar K^{*0})=(-0.7\pm5.0_{\rm stat.} \pm4.1_{\rm syst.})\%$ is consistent with  $C\!P$ conservation.

\end{abstract}


\maketitle


\oddsidemargin  -0.2cm
\evensidemargin -0.2cm

\section{Introduction}

Charmed meson decays provide a sensitive laboratory for probing non-perturbative Quantum Chromodynamics (QCD). 
The radiative decays $D^0 \to \gamma V$ ($V$ denotes light vector mesons) receive contributions from both short-distance and long-distance interactions. 
Although the short-distance amplitude is enhanced by QCD corrections, it remains smaller than the long-distance contribution~\cite{ref9}. 
The latter, dominated by non-perturbative mechanisms such as those described by the Vector Meson Dominance (VMD) model, can enhance branching fractions (BFs) by up to $10^{-4}$.

Predictions from various theoretical approaches for the $D^0\to \gamma \bar K^{*0}$ [$\bar K^{*0}$ denotes $\bar K^{*}(892)^0$] and $D^0\to \gamma \phi$ BFs are summarized in Table~\ref{theory}. 
The decay $D^0\to\gamma \phi$ was first observed by the Belle collaboration~\cite{ref3} and subsequently confirmed by BaBar with improved precision~\cite{ref4}. 
The decay $D^0\to\gamma \bar K^{*0}$ was first observed by BaBar~\cite{ref4} and later confirmed by Belle~\cite{ref6}, both of which were relative measurements. In contrast, this work presents the first absolute measurement of this decay. The corresponding measurements are listed in Table~\ref{theory}. 

Notably, the latest Belle~\cite{ref6} measurement of $D^0\to\gamma \bar K^{*0}$ differs from the BaBar result by about $3\sigma$. 
Improved measurements of radiative $D^0$ decays are therefore essential to clarify this tension and to further constrain the underlying decay dynamics. 
Throughout this paper, charge-conjugated decays are implied unless otherwise stated.

Measurements of charge-conjugation and parity ($C\!P$) asymmetries in radiative charm decays provide complementary information on weak decay mechanisms in the charm sector. 
Within the Standard Model, direct $C\!P$ violation in charm decays is expected to be at most of order $10^{-3}$. 
Evidence for $C\!P$ violation at the $10^{-4}$ level has been reported by LHCb in $D^0\to K^+K^-$ and $D^0\to \pi^+\pi^-$ decays~\cite{cp}. 
The Belle collaboration~\cite{ref6} reported the first measurement of the $C\!P$ asymmetry in radiative charm decays, with results consistent with $C\!P$ conservation. 

In this paper, we present measurements of the absolute BFs of the decays $D^0\to\gamma \bar K^{*0}$ and $D^0\to\gamma \phi$, as well as the asymmetry $\mathcal{A}_{C\!P}(D^0\to\gamma \bar K^{*0})$, using 20.3 fb$^{-1}$ of $e^+e^-$ collision data~\cite{lum_bes3,lum_bes31} collected with the BESIII detector at a center-of-mass energy of $\sqrt{s}=3.773$~GeV. 
This energy is near the $\psi(3770)$ resonance, which predominantly decays into $D\bar D$ ($D$ denotes $D^0$ or $D^+$) pairs. 
Operating close to the $D\bar D$ production threshold, the $D$ mesons are produced nearly at rest, providing a clean kinematic environment for studying their decays.
			
\begin{table}[htbp]
\centering
\caption{\small  Theoretical calculations and  experimental measurements for the BFs~ (in unit of $\times10^{-5}$) of $D^0\to \gamma \bar K^{*0} $ and $D^0\to \gamma \phi$. The theoretical calculations include: Vector Meson Dominance (VMD), Heavy Quark Effective Theory (HQET), Effective Lagrangian Approach (ELA), Non-Perturbative QCD (NPQCD), Effective Field Theory (EFT), and Left-Right Symmetric Model (LRSM).	
}
\small
				
\begin{tabular}{lcc}
  \hline
  \hline 
  Theoretical calculation   &${D^0 \to \gamma \bar K^{*0}}$& ${D^{0}\to \gamma \phi}$ \\
  \hline
    VMD 1~\cite{ref9}&7-12& 0.1-3.4 \\
     VMD 2~\cite{ref11}&       16-34    &- \\ 
    HQET 1~\cite{ref7}& 21-43 &- \\
 HQET 2~\cite{ref8}&20-40 &- \\
   HQET 3~\cite{ref99}& 2.6-46    &0.24-2.8 \\ 
     HQET 4~\cite{ref13}& $\leq$3.0 &2.0 \\ 
   HQET 5~\cite{ref14}& 7-20 &0.4-3 \\ 
   ELA~\cite{ref10}& 11&- \\
   NPQCD~\cite{{ref88}}& $3.59^{+0.20}_{-0.22}$ &$2.76^{+0.36}_{-0.35}$ \\ 
    EFT~\cite{ref12}& 6-36 &0.1-1.9 \\ 
  LRSM~\cite{Biswas:2017eyn}& 4.6-18 &0.48-0.64 \\ 
 
  \hline
  \hline 
   Experimental measurement  &${D^0 \to \gamma \bar K^{*0}}$& ${D^{0}\to \gamma \phi}$ \\
  \hline
 BaBar~\cite{ref4}& $33.3\pm2.0\pm2.8$ &$2.82\pm0.31\pm0.27$ \\
 Belle~\cite{ref6}& $46.9\pm2.0\pm2.0$ &$2.71\pm0.19\pm0.09$ \\
  \hline
  \hline
\end{tabular}

\label{theory}
\end{table}

\section{BESIII detector and Monte Carlo}

The BESIII detector~\cite{BESIII} records symmetric $e^+e^-$ collisions 
provided by the BEPCII storage ring~\cite{Yu:IPAC2016-TUYA01}
in the center-of-mass energy range from $\sqrt{s}=1.84$ to $4.95$~GeV,
with a peak luminosity of $1.1 \times 10^{33}\;\text{cm}^{-2}\text{s}^{-1}$ achieved at $\sqrt{s} = 3.773\;\text{GeV}$. 
BESIII has collected large data samples in this energy region~\cite{cpc41,EcmsMea,EventFilter}. 
The cylindrical core of the BESIII detector covers 93\% of the full solid angle and consists of a helium-based multilayer drift chamber (MDC), a plastic scintillator time-of-flight system (TOF), and a CsI(Tl) electromagnetic calorimeter (EMC), all enclosed in a superconducting solenoidal magnet providing a 1.0~T magnetic field. 
The solenoid is supported by an octagonal flux-return yoke with resistive plate counter muon identifier modules interleaved with steel.  
The charged-particle momentum resolution at $1~{\rm GeV}/c$ is $0.5\%$, and the specific ionization energy loss ${\rm d}E/{\rm d}x$ resolution is $6\%$ for electrons from Bhabha scattering. 
The EMC measures photon energies with a resolution of $2.5\%$ ($5\%$) at $1$~GeV in the barrel (end cap) region. 
The time resolution of the TOF barrel section is 68~ps, while that of the end cap section was 110~ps. 
The end cap TOF system was upgraded in 2015 using multigap resistive plate chamber technology, providing a time resolution of 60~ps, which benefits 85.5\% of the data used in this analysis~\cite{updatadata}.  
  
Simulated samples produced with the {\sc geant4}-based~\cite{geant4} Monte Carlo (MC) package, which includes the geometric description of the BESIII detector and its response, are used to determine the detection efficiency and estimate backgrounds. 
The simulation includes the beam-energy spread and initial-state radiation (ISR) in $e^+e^-$ annihilation modeled with the generator {\sc kkmc}~\cite{kkmc}. 
The inclusive MC samples consist of quantum-coherent neutral $D\bar{D}$ production, non-$D\bar{D}$ decays of the $\psi(3770)$, ISR production of the $J/\psi$ and $\psi(3686)$ states, and continuum processes. 
Known decay modes are modeled with {\sc evtgen}~\cite{evtgen1,evtgen2} using BFs taken from the Particle Data Group (PDG)~\cite{pdg2022}, while the remaining unknown charmonium decays are modeled with {\sc lundcharm}~\cite{lundcharm1,lundcharm2}. 
Final-state radiation from charged particles is simulated with the {\sc photos} package~\cite{photos}.

\section{Measurement method}

This analysis is performed using the double-tag (DT) method, which was first developed at the Mark~III experiment~\cite{mark3}.
If a $\bar D^0$ meson is fully reconstructed in one of the following six tag modes: $\bar D^0\to K^+\pi^-$, $\bar D^0\to K^+\pi^-\pi^0$, $\bar D^0\to K^{+}\pi^{-}\pi^{-}\pi^{+}$, $\bar D^0\to K^{+}\pi^{-}\pi^{-}\pi^{+}\pi^0$, $\bar D^0\to K^{+}\pi^{-}\pi^0\pi^0$, and $\bar D^0\to K^0_S\pi^{+}\pi^{-}$, it is referred to as a single-tag (ST) candidate. A DT event~\cite{ke2023} is defined as an event containing either a signal decay of $D^0\to\gamma \bar{K}^{*0}$ or $D^0\to\gamma\phi$, in conjunction with an ST $\bar{D}^0$. Consequently,
the BF of the signal decay is determined by
\begin{equation}
\label{eq:br}
{\mathcal B}_{{\rm sig}} = \frac{N_{\rm DT}} {N^{\rm tot}_{\rm ST} \cdot {\epsilon_{\rm sig}} \cdot {\mathcal B}_{\rm sub}},
\end{equation}
where $N_{\rm DT}$ is the number of events with any $\bar D^0$ tag and a signal candidate. 
$N^{\rm tot}_{\rm ST} = {\sum\limits_{i=1}^{6} N_{\rm ST}^i}$ is the total ST yield, $N_{\rm ST}^i$ is the number of ST $\bar D^0$ candidates in tag mode $i$,
$\epsilon_{\rm sig}=\sum\limits_{i=1}^{6}[(N_{\rm ST}^i\cdot \epsilon_{\rm DT}^i)/(N^{\rm tot}_{\rm ST}\cdot \epsilon_{\rm ST}^i)]$ is the signal efficiency in the presence of the ST candidate,
where $\epsilon_{\rm ST}^{i}$ and $\epsilon_{\rm DT}^{i}$ are the efficiencies for selecting ST and DT candidates, respectively, $i$ denotes the tag modes. ${\mathcal B}_{\rm sub}$ is the BF of $\bar K^{*0}\to K^-\pi^+$ or $\phi \to K^+K^-$.

\subsection{Single-Tag Selection}

\begin{figure*}[htp]
  \centering
\includegraphics[width=0.85\linewidth]{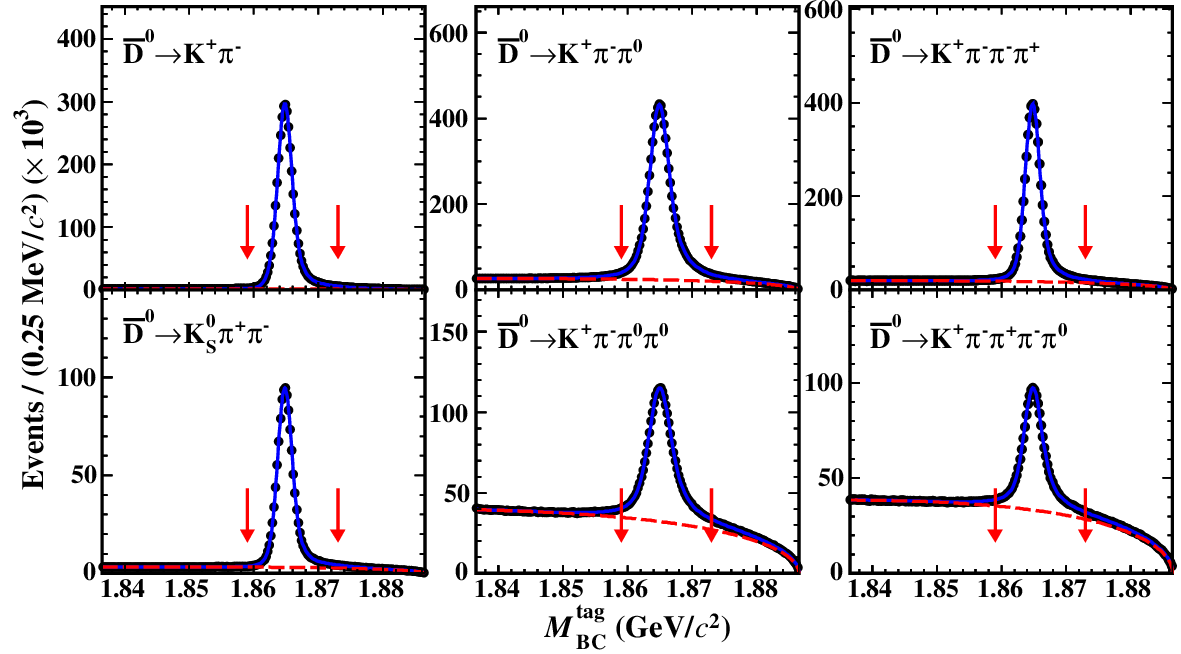}
  \caption{\small
Fits to the $M_{\rm BC}^{\rm tag}$ distributions of the ST $\bar D^0$ candidates. Points with error bars represent the data; the solid curve shows the total fit, and the dashed curve shows the background component. The pairs of arrows denote the chosen signal windows.}
\label{fig:STfit}
\end{figure*}

Candidates for $K^0_S$ and $\pi^0$ are reconstructed via $K^0_S\to\pi^+\pi^-$ and $\pi^0\to\gamma\gamma$, respectively.
Candidates for $K^\pm$, $\pi^\pm$,
$K^0_S$, $\gamma$, and $\pi^0$ are reconstructed and identified using the same criteria as those in
Refs.~\cite{cpc40,papernew1,kkpipi,DCS-kpipi}.
The tagged $\bar D^0$ mesons are selected using two variables: the energy difference
\begin{equation}\label{def_delE}
\Delta E_{\rm tag} \equiv E_{\bar D^0} - E_{\rm b},
\end{equation}
and the beam-energy-constrained mass
\begin{equation}\label{def_mbc}
M_{\rm BC}^{\rm tag} \equiv \sqrt{E^{2}_{\rm b}/c^4-|\vec{p}_{\bar D^0}|^{2}/c^2},
\end{equation}
where $E_{\rm b}$ is the beam energy, $\vec{p}_{\bar D^0}$ and $E_{\bar D^0}$ are the momentum and energy of the $\bar D^0$ candidate in the $e^+e^-$ rest frame, respectively.
If multiple combinations are found in an event, only the combination with the minimum $|\Delta E_{\rm tag}|$ is retained for further analysis.
To suppress  combinatorial backgrounds in the $M_{\rm BC}$ distribution, 
tag dependent $\Delta E$ requirements are imposed on the ST candidates. 
The detailed $\Delta E$ requirements  are summarized in Table~\ref{tab:ST}.
The yields of ST $\bar D^0$ mesons are obtained from unbinned maximum-likelihood fits to the $M_{\rm BC}^{\rm tag}$ distributions of the accepted ST candidates.
In the fit, the signal is described by the MC-simulated shape, requiring the opening angle between the generated and reconstructed three-momenta of ST daughter particles to be less than $15^\circ$, and convolved with a double Gaussian function plus an additional single Gaussian. The combinatorial background is described by an ARGUS function~\cite{ARGUS}.
The parameters of the ARGUS function are floated, except for the endpoint, which is fixed at 1.8865~$\mathrm{GeV}/c^2$. The fit results are shown in Fig.~\ref{fig:STfit}. Candidates with $M_{\rm BC}^{\rm tag}\in (1.859,\,1.873)$ GeV/$c^2$ are retained for further analysis. The ST yields in data and the ST efficiencies for different tag modes are summarized in Table~\ref{tab:ST}. For ST candidates of $\bar D^0\to K^+\pi^-$, backgrounds from cosmic rays and Bhabha events are further suppressed using the requirement described in Ref.~\cite{yuzhouxian}.

\begin{table}[htp]
\centering
\caption{The $\Delta E$ requirements, ST yields in data, and the ST efficiencies for different tag modes, where the uncertainties are statistical, and the efficiencies do not include the BFs of $K^0_S\to \pi^+\pi^-$ and $\pi^0\to \gamma\gamma$.
}
\label{tab:ST}
\centering
				\resizebox{\linewidth}{!}{

\begin{tabular}{lccc}
  \hline\hline
$\bar D^0$ mode&$\Delta E (\rm GeV)$&$N_{\rm ST}^i~(\times 10^{3})$& $\epsilon^{i}_{\rm ST}$\,(\%) \\ \hline
$K^{+}\pi^{-}$             & $(-0.027,\,0.027)$      &  $3725.7\pm2.0$       &$65.10\pm0.01$\\
$K^{+}\pi^{-}\pi^{0}$      & $(-0.062,\,0.049)$        &  $7422.3\pm3.2$     &$35.60\pm0.01$\\
$K^{+}\pi^{-}\pi^{-}\pi^{+}$    & $(-0.026,\,0.024)$   &  $4987.5\pm2.5$      &$40.94\pm0.01$\\
$K^{+}\pi^{-}\pi^{-}\pi^{+}\pi^0$& $(-0.057,\,0.051)$ &  $1154.9\pm1.8$      &$16.27\pm0.01$\\
$K^{+}\pi^{-}\pi^0\pi^0$    & $(-0.068,\,0.053)$      &  $1770.9\pm2.2$      &$15.11\pm0.01$\\
$K^0_S\pi^{+}\pi^{-}$      & $(-0.024,\,0.024)$      &  $1168.6\pm1.2$        &$37.90\pm0.01$\\
\hline
$N^{\rm tot}_{\rm ST}$             &-    &20229.9 $\pm$ 5.5  &-\\
\hline\hline
\end{tabular}
}
\end{table}

\subsection{Signal Selection}

To form candidates for $D^0\to \gamma \bar K^{*0}$ and $D^0\to \gamma \phi$, the $\bar K^{*0}$ and $\phi$ are reconstructed via their decays $\bar K^{*0}\to K^-\pi^+$ and $\phi\to K^+K^-$. The $\gamma$, $K^\pm$, and $\pi^\pm$ are selected using the same criteria as on the tag side. It is required that no extra charged tracks are present to suppress background from hadronic $D$ decays. The most energetic $\gamma$ is retained as the radiative photon. 

The invariant masses $M_{K^-\pi^+}$ and $M_{K^+K^-}$ of the $\bar K^{*0}$ and $\phi$ candidates are required to satisfy
$M_{K^-\pi^+}\in (0.855,\,0.945)$~GeV/$c^2$ and $M_{K^+K^-}\in (1.010,\,1.030)$~GeV/$c^2$. 
The former requirement is determined by optimizing the figure of merit $S/\sqrt{S+B}$, where $S$ and $B$ are the signal and background yields in the inclusive MC sample, respectively. The latter corresponds to approximately twice the $\phi$ width around the mass given by the PDG~\cite{pdg2022}. 
To further suppress backgrounds associated with extra $\pi^0$ mesons, events containing any additional two-photon combination that satisfies the $\pi^0$ selection criteria are rejected.

\begin{figure}[htp]
  \centering
\includegraphics[width=0.6\linewidth]{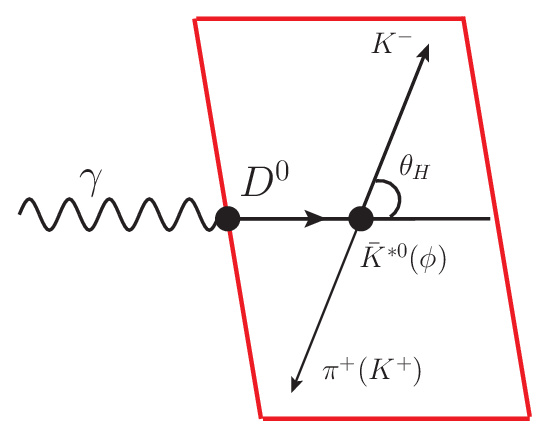}
  \caption{
Definition of the helicity angle of $K^-$ diagram in the rest frame of $\bar K^{*0}$ or $\phi$. The horizontal arrow denotes the direction of $\bar K^{*0}$, and the slanted arrow denotes the direction of $K^{-}$.}
\label{plot:Helix}
\end{figure}

On the signal side, the energy difference and the beam-energy-constrained mass, $\Delta E_{\rm sig}$ and $M_{\rm BC}^{\rm sig}$, are defined analogously to those on the tag side. To suppress combinatorial backgrounds, we require $M_{\rm BC}^{\rm sig}\in (1.859,\,1.873)$~GeV/$c^2$. 
To further discriminate signal from background, the helicity angle $\theta_H$ of the $K^-$ is defined in the rest frame of the $\bar K^{*0}$ or $\phi$, as shown in Fig.~\ref{plot:Helix}. 
Due to angular-momentum conservation, the signal is expected to follow a $1-\cos^2\theta_H$ distribution.

\section{Branching Fraction}

\begin{figure*}[htp]
  \centering
  \includegraphics[width=0.85\linewidth]{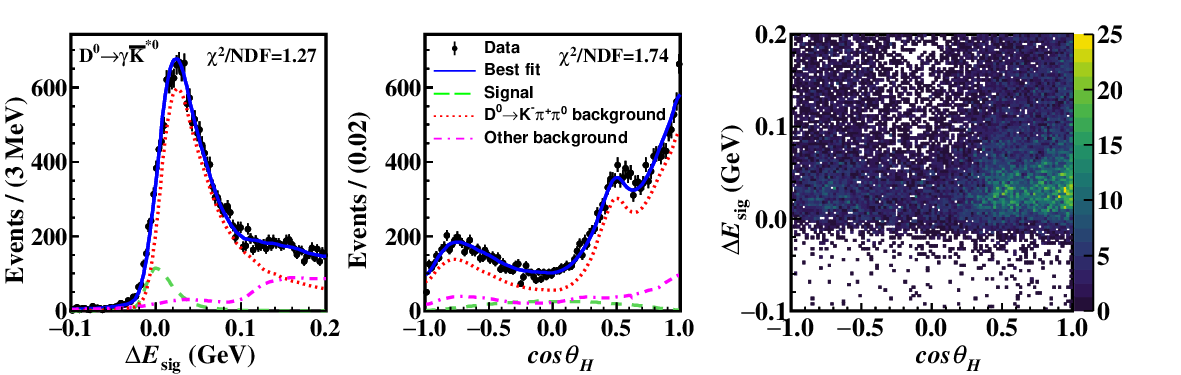}
  \includegraphics[width=0.85\linewidth]{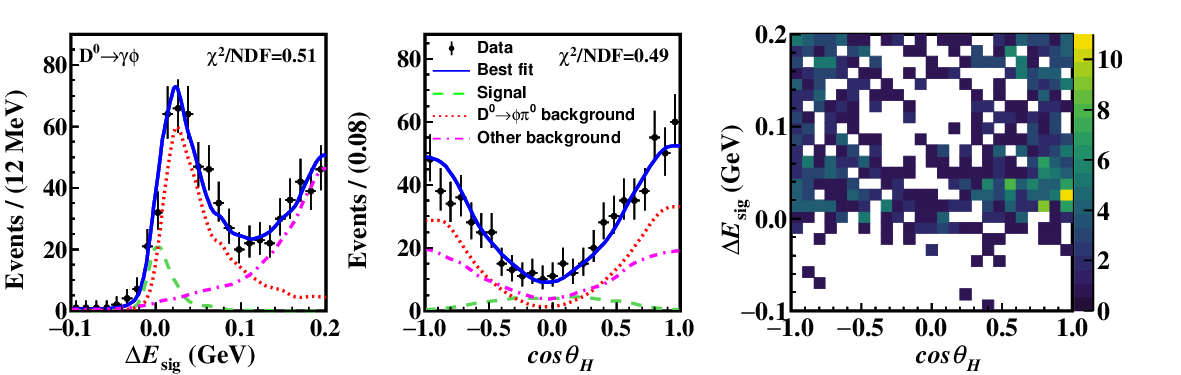}
  \caption{
Projections onto $\Delta E_{\rm sig}$ (left column) and $\cos \theta_{H}$ (middle column) of the 2D fits and the 2D distributions (right column) for $D^0\to\gamma \bar K^{*0}$ (top row) and $D^0\to\gamma \phi$ (bottom row) candidate events in data. The points with error bars are data. The blue solid curves are the fit results. The green dashed curves are the fitted signal shapes, the red dotted curves are the $D^0\to K^-\pi^+\pi^0$ or $D^0\to \phi \pi^0$ backgrounds,  and the pink dash-dotted curves are the other processes. The events in the  2D distributions are data.
}
\label{2Dfit1}
\end{figure*}

To extract the DT yields, a two-dimensional (2D) extended unbinned maximum-likelihood fit
is performed to the $\Delta E_{\rm sig}$ versus $\cos\theta_H$ distribution.
The 2D kernel estimation probability density function (KeysPDF) for the signal is obtained from the MC simulation and modeled using the signal MC shape.
The dominant backgrounds are $D^0 \to K^-\pi^+\pi^0$ (amplitude analysis model, following the method of Ref.~\cite{AMP}) and $D^0 \to \phi\pi^0$ (SVS model: scalar to a vector and a scalar meson~\cite{evtgen2,svs}) for $D^0\to\gamma \bar K^{*0}$ and $D^0\to\gamma \phi$, respectively. 
The background distributions as functions of $\Delta E_{\rm sig}$ and the helicity angle $\theta_{H}$ are obtained from the inclusive MC sample.
All yields are allowed to float in the fits.
The spectra in the left and middle columns of Fig.~\ref{2Dfit1} show the projections onto $\Delta E_{\rm sig}$ and $\cos \theta_{H}$ from the 2D fits to data. From the fits, the signal yields for $D^0\to\gamma \bar K^{*0}$ and
$D^0\to\gamma \phi$ are determined to be $1652.9\pm77.0$ and $69.1\pm12.0$ events, respectively, where the uncertainties are statistical only.
For each signal decay, the statistical significance is evaluated by comparing the likelihood values from fits with and without the signal component, accounting for the change in degrees of freedom (NDF). The statistical significances for $D^0\to\gamma \bar K^{*0}$ and $D^0\to\gamma \phi$ are $26.8\sigma$ and $7.9\sigma$, respectively. The BFs for the decays
$D^0\to\gamma \bar K^{*0}$ and $D^0\to\gamma\phi$ are measured to be $(3.81 \pm 0.18_{\rm stat.} \pm 0.20_{\rm syst.})\times 10^{-4}$ and $(2.51 \pm 0.44_{\rm stat.} \pm 0.11_{\rm syst.})\times 10^{-5}$, respectively. The systematic uncertainties are discussed in the next section.
The obtained values of ${ N_{{\rm DT}}}$, $\epsilon_{{\rm sig}}$, and $\mathcal B_{\rm sig}$ are summarized in Table~\ref{tab:DT}.

\begin{figure*}[]
	\centering
	\includegraphics[width=0.75\linewidth]{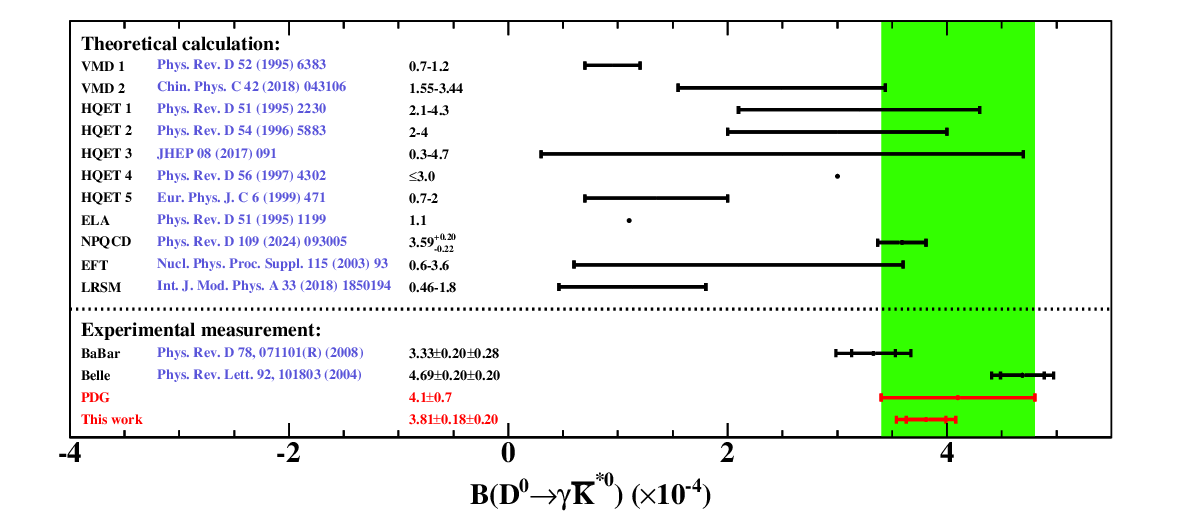}
	\includegraphics[width=0.75\linewidth]{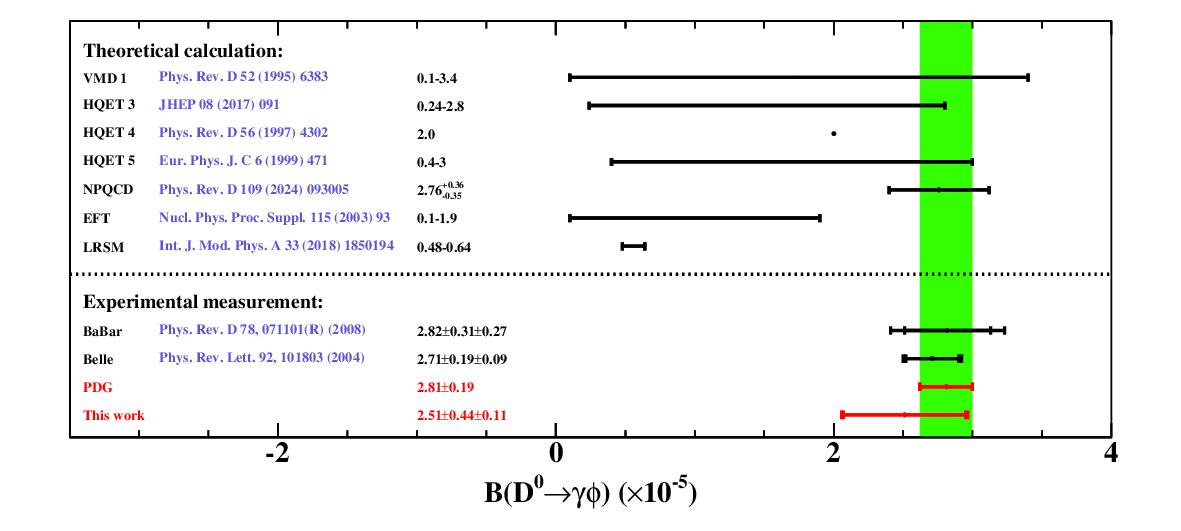}
	\caption{
		Comparisons of the obtained BFs with theoretical calculations,  other measurements, and the world average values (shown as green bands). For the measurements of BESIII, Belle, and BaBar, the inner error bars are statistical uncertainties, and the outer error bars are the combined statistical and systematic uncertainties in quadrature. }
	\label{fit:20fbbfcompare}
\end{figure*}

  Figure~\ref{fit:20fbbfcompare} compares the measured BFs with previous measurements and the world average values. 
The BF of $D^0 \to \gamma \bar K^{*0}$ measured in this work is consistent with the predictions of the VMD2~\cite{ref11}, HQET~\cite{ref7,ref8,ref99,ref13,ref14}, NPQCD~\cite{ref88}, and EFT~\cite{ref12} models, while it deviates from those of the VMD1~\cite{ref9}, ELA~\cite{ref10}, and LRSM~\cite{Biswas:2017eyn} models. 
The BF of $D^0\to \gamma \phi$ reported here is consistent with the predictions of the VMD1~\cite{ref9}, HQET~\cite{ref99,ref13,ref14}, NPQCD~\cite{ref88}, and EFT~\cite{ref12} models, and deviates from that of the LRSM model~\cite{Biswas:2017eyn}.

\begin{table*}[htp]
\centering
\caption{\label{tab:DT}
The DT yields in data ($N_{\rm DT}$), statistical significances, signal efficiencies ($\epsilon_{\rm sig}$) and  $\mathcal B_{\rm sig}$(in unit of $\times 10^{-5}$) .
The efficiencies do not include the BFs of $\bar K^{*0}$ and $\phi$ decays. The first uncertainties are statistical and the second systematic.}
\begin{tabular}{lcccccc}
\hline\hline
Decay mode   & $N_{\rm DT}$ &Significance & $\epsilon_{\rm sig}$\,(\%)& $\mathcal B_{\rm sig}$ (This work)&$\mathcal B_{\rm sig}$ (Belle)~\cite{ref6}&$\mathcal B_{\rm sig}$ (BaBar)~\cite{ref4}\\ \hline
$D^0\to \gamma \bar K^{*0}$ &$1652.9\pm77.0$&$26.8\sigma$&$32.17\pm0.07$&$38.1\pm1.8\pm2.0$&$46.9\pm2.0\pm2.0$&$33.3\pm2.0\pm2.8$ \\
$D^0\to \gamma \phi$        &$69.1\pm12.0$&$7.9\sigma$&$27.29\pm0.07$&$2.51\pm0.44\pm0.11$&$2.71\pm0.19\pm0.09$&$2.83\pm0.31\pm0.27$\\
\hline\hline
\end{tabular}
\end{table*}

\section{Systematic Uncertainties}

By exploiting the DT method, most uncertainties related to the ST selection cancel.
The systematic uncertainties in the BF measurements arise from the following sources and are assigned relative to the measured BFs.

The systematic uncertainty on the total ST $\bar D^0$ yield is estimated by varying the signal and background shapes. For the background, the endpoint of the ARGUS function is varied, resulting in an uncertainty of $0.02\%$. For the signal, the mean values, widths, and correlation coefficients of the three Gaussian functions are varied by $1\sigma$, yielding an uncertainty of $0.25\%$. The total uncertainty from this source is taken as $0.3\%$.

The tag bias arising from different reconstruction environments of the tag side in the inclusive and signal MC samples is assigned as $0.2\%$ for $D^0\to\gamma \bar K^{*0}$ and $0.4\%$ for $D^0\to\gamma \phi$.

The tracking and particle identification (PID) efficiencies of $K^\pm$ and $\pi^\pm$ are studied using DT $D\bar D$ hadronic events~\cite{bes3-pimuv}. A systematic uncertainty of $0.5\%$ per $K^\pm$ and $\pi^\pm$ is assigned for tracking and PID. The systematic uncertainty associated with photon selection is taken as $1.0\%$, determined using $J/\psi\to \pi^+\pi^-\pi^0$ events~\cite{gammaselection}.

The systematic uncertainty associated with the $M_{\rm BC}^{\rm sig}$ requirement is evaluated using control samples of $D^0\to K^-\pi^+\pi^0$ and $D^0\to K^-\pi^+$. The largest difference between data and the inclusive MC sample, $0.5\%$, is assigned as the systematic uncertainty.

The systematic uncertainties associated with the $\bar K^{*0}$ and $\phi$ mass windows are determined using control samples of $D^+\to \bar K^{*0}(\to K^-\pi^+)e^+\nu_e$ and $D^+\to \phi(\to K^+K^-)\pi^+$. The $M_{K^-\pi^+}$ and $M_{K^+K^-}$ invariant mass spectra are fitted to the data and the MC simulation. The signal is modeled with a Breit-Wigner function and the background with a polynomial function. By comparing the yields inside and outside the signal mass regions, the differences in acceptance efficiencies between data and MC are determined to be $2.5\%$ and $1.9\%$, respectively, and are assigned as systematic uncertainties.

The systematic uncertainties due to the requirements of no extra $\pi^0$ and no extra charged tracks are evaluated using $D^0\to K^-\pi^+\pi^0$ and $D^0\to K^-\pi^+$ candidates. The largest differences in acceptance efficiencies between data and MC simulation, $1.3\%$ and $1.6\%$, are assigned as the systematic uncertainties for the no-extra-$\pi^0$ and no-extra-track requirements, respectively.

The systematic uncertainty associated with the 2D fit is estimated by repeating the BF measurements using alternative signal shapes obtained by smearing with a Gaussian resolution function, varying the Gaussian resolution and the smoothness parameters of the KeysPDF, and changing the fit ranges. For $D^0\to \gamma \bar K^{*0}$, the maximum variation of $3.7\%$ in the BF is taken as the systematic uncertainty; for $D^0\to \gamma \phi$, it is $2.7\%$.

The limited MC sample size contributes a $0.3\%$ systematic uncertainty for both $D^0 \to \gamma \bar K^{*0}$ and $D^0 \to \gamma \phi$. The quoted BF of $\phi\to K^+K^-$ is $(49.9\pm0.5)\%$~\cite{pdg2022}, corresponding to a $1.0\%$ uncertainty.

Adding all contributions in quadrature, the total systematic uncertainties are $5.3\%$ for $D^0\to \gamma \bar K^{*0}$ and $4.4\%$ for $D^0\to \gamma \phi$.

\begin{table}[]
\centering
\caption{Relative systematic uncertainties (\%) in the BF measurements. The systematic uncertainties marked with an asterisk are independent of the $C\!P$ asymmetry measurement.}
\begin{tabular}{lcc}
  \hline
  \hline 
  Source  &${D^0 \to \gamma \bar K^{*0}}$& ${D^{0}\to \gamma \phi}$ \\
  \hline
$N_{\rm tag}$                      &$0.3$&0.3\\
Tag bias                           &$0.2^{*}$&0.4\\
Tracking                           &$1.0^{*}$&1.0\\
PID                                &$1.0^{*}$&1.0\\
$\gamma$ selection                 &1.0&1.0\\
$M_{\rm BC}^{\rm sig}$ requirement &0.5&0.5\\
$\bar K^{*0}$ or $\phi$ mass window&2.5&1.9\\
$\pi^0$ veto                       &$1.3$&1.3\\
2D fit                             &$3.7^{*}$&2.7\\
Quoted BF          &-&1.0\\
MC statistics                      &$0.3^{*}$&0.3\\
$N_{\rm charge}^{\rm extra}$               &$1.6^{*}$&1.6\\
\hline
Total                              &5.3&4.4\\
  \hline
  \hline
\end{tabular}
\label{sys1}
\end{table}

\section{CP Asymmetry Measurement}

Furthermore, the BFs of $D^0\to \gamma \bar K^{*0}$ and $\bar D^0\to \gamma K^{*0}$, denoted as $\mathcal B^{+}$ and $\mathcal B^{-}$, are measured separately to determine the ${C\!P}$ asymmetry, defined as 
$\mathcal{A}_{C\!P}=\frac{{\mathcal B^{+}}-{\mathcal B^{-}}}{{\mathcal B^{+}}+{\mathcal B^{-}}}$.
The $K_{S}^{0}\pi^{+}\pi^{-}$ ST channel is excluded because it cannot distinguish between $D^{0}$ and $\bar D^{0}$ mesons. 
Using the same method described above, we obtain DT yields, BFs, and asymmetry, which are summarized in Table~\ref{cp}.

In the determination of $\mathcal{A}_{C\!P}$, the systematic uncertainties associated with $\gamma$ selection, $\pi^{0}$ veto, $N_{\rm tag}$, the $M_{\rm BC}$ requirement, and the $\bar K^{*0}$ mass window are treated as correlated between charge-conjugated modes, while the remaining uncertainties are treated as uncorrelated. 
The contributions are Tag bias ($0.4\%$, $0.4\%$), Tracking ($0.1\%$, $0.1\%$), PID ($0.1\%$, $0.1\%$), 2D fit ($5.6\%$, $5.3\%$), MC statistics ($0.3\%$, $0.3\%$), and $N^{\rm extra}_{\rm charge}$ ($1.7\%$, $1.7\%$), where the first value in each pair corresponds to $D^0\to \gamma \bar K^{*0}$ and the second to $\bar D^0\to \gamma K^{*0}$. 
The total systematic uncertainties, evaluated separately for the two charge states using the same procedure as described above, are ($5.9\%$, $5.6\%$).

The resulting ${C\!P}$ asymmetry is 
$\mathcal{A}_{C\!P}=(-0.7\pm5.0_{\rm stat.}\pm4.1_{\rm syst.})\%$, 
which is consistent with zero and indicates no evidence for ${C\!P}$ violation. 
Our result is consistent with the Belle measurement~\cite{ref6}.

\begin{table}[h]
	\centering
	\caption{\label{cp}
Signal yields ($N_{\rm net}$), charge-separated BFs ($\mathcal B^{+}$ and $\mathcal B^{-}$), and their asymmetries ($\mathcal{A}_{C\!P}$). The first and second uncertainties are statistical and systematic, respectively. }
	\resizebox{\linewidth}{!}{
		\begin{tabular}{lcccc}
			\hline\hline
			Decay mode   &$\mathcal \rm N_{\rm net}$  &$\mathcal B ~(\times 10^{-4})$ & $\mathcal{A}_{C\!P}$ (\%)&$\mathcal{A}_{C\!P}$ (Belle) $(\%)$\\ \hline
			$D^0\to \gamma \bar K^{*0}$ &$754.4\pm53.4$&$3.72\pm0.26\pm0.20$&\multirow{2}{*}{$-0.7\pm5.0\pm4.1$}&\multirow{2}{*}{$-0.3\pm2.0\pm0.0$}\\
			$\bar D^0\to \gamma K^{*0}$ &$769.6\pm53.2$&$3.77\pm0.26\pm0.21$ \\			
			\hline\hline
		\end{tabular}
	}
\end{table}

\section{Summary}

Using a data sample of $20.3\,\mathrm{fb}^{-1}$ collected at $\sqrt{s}=3.773$\,GeV with the BESIII detector,
the absolute BFs of the radiative decays $D^0 \to \gamma \bar K^{*0}$ and $D^0\to \gamma \phi$ are measured for the first time, yielding $(3.81 \pm 0.18_{\rm stat.} \pm 0.20_{\rm syst.})\times 10^{-4}$ and $(2.51 \pm 0.44_{\rm stat.} \pm 0.11_{\rm syst.})\times 10^{-5}$, respectively. The measured value is consistent with the BaBar and Belle results within $1.1\sigma$ and $2.0\sigma$, respectively, and achieves improved precision compared to the PDG average~\cite{pdg2022}. 
In addition, the $C\!P$ asymmetry between $D^0\to \gamma \bar K^{*0}$ and $\bar D^0\to \gamma K^{*0}$ is measured to be 
$\mathcal{A}_{C\!P}(D^0\to\gamma \bar K^{*0}) = (-0.7\pm5.0_{\rm stat.}\pm4.1_{\rm syst.})\%$, 
which is consistent with $C\!P$ conservation. 
Future experiments such as Belle~II~\cite{bellebaipi} and STCF~\cite{STCF1,STCF2} are expected to provide more precise measurements of radiative charm decays.
 
\section{Acknowledgement}

The BESIII Collaboration thanks the staff of BEPCII (https://cstr.cn/31109.02.BEPC) and the IHEP computing center for their strong support. This work is supported in part by National Key R\&D Program of China under Contracts Nos. 2023YFA1606000, 2023YFA1606704, 2025YFA1613900; National Natural Science Foundation of China (NSFC) under Contracts Nos. 11635010, 11935015, 11935016, 11935018, 12025502, 12035009, 12035013, 12061131003, 12192260, 12192261, 12192262, 12192263, 12192264, 12192265, 12221005, 12225509, 12235017, 12342502, 12361141819; the Chinese Academy of Sciences (CAS) Large-Scale Scientific Facility Program; the Strategic Priority Research Program of Chinese Academy of Sciences under Contract No. XDA0480600; CAS under Contract No. YSBR-101; 100 Talents Program of CAS; The Institute of Nuclear and Particle Physics (INPAC) and Shanghai Key Laboratory for Particle Physics and Cosmology; ERC under Contract No. 758462; German Research Foundation DFG under Contract No. FOR5327; Istituto Nazionale di Fisica Nucleare, Italy; Knut and Alice Wallenberg Foundation under Contracts Nos. 2021.0174, 2021.0299, 2023.0315; Ministry of Development of Turkey under Contract No. DPT2006K-120470; National Research Foundation of Korea under Contract No. NRF-2022R1A2C1092335; National Science and Technology fund of Mongolia; Polish National Science Centre under Contract No. 2024/53/B/ST2/00975; STFC (United Kingdom); Swedish Research Council under Contract No. 2019.04595; U. S. Department of Energy under Contract No. DE-FG02-05ER41374

\author{M.~Ablikim$^{1}$\BESIIIorcid{0000-0002-3935-619X},
	M.~N.~Achasov$^{4,c}$\BESIIIorcid{0000-0002-9400-8622},
	P.~Adlarson$^{83}$\BESIIIorcid{0000-0001-6280-3851},
	X.~C.~Ai$^{89}$\BESIIIorcid{0000-0003-3856-2415},
	C.~S.~Akondi$^{31A,31B}$\BESIIIorcid{0000-0001-6303-5217},
	R.~Aliberti$^{39}$\BESIIIorcid{0000-0003-3500-4012},
	A.~Amoroso$^{82A,82C}$\BESIIIorcid{0000-0002-3095-8610},
	Q.~An$^{79,65,\dagger}$,
	Y.~H.~An$^{89}$\BESIIIorcid{0009-0008-3419-0849},
	Y.~Bai$^{63}$\BESIIIorcid{0000-0001-6593-5665}
% ...
	\\
	\vspace{0.2cm}
	(BESIII Collaboration)\\
	\vspace{0.2cm} {\it
		$^{1}$ Institute of High Energy Physics, Beijing 100049, People's Republic of China\\
% ...
$^{4}$ Budker Institute of Nuclear Physics SB RAS (BINP), Novosibirsk 630090, Russia\\
% ...
$^{31}$ INFN Sezione di Ferrara, (A)INFN Sezione di Ferrara, I-44122, Ferrara, Italy; (B)University of Ferrara, I-44122, Ferrara, Italy\\
% ...
$^{39}$ Johannes Gutenberg University of Mainz, Johann-Joachim-Becher-Weg 45, D-55099 Mainz, Germany\\
% ...
$^{63}$ Southeast University, Nanjing 211100, People's Republic of China\\
% ...
$^{65}$ State Key Laboratory of Particle Detection and Electronics, Beijing 100049, Hefei 230026, People's Republic of China\\
% ...
$^{79}$ University of Science and Technology of China, Hefei 230026, People's Republic of China\\
% ...
$^{82}$ University of Turin and INFN, (A)University of Turin, I-10125, Turin, Italy; (B)University of Eastern Piedmont, I-15121, Alessandria, Italy; (C)INFN, I-10125, Turin, Italy\\
		$^{83}$ Uppsala University, Box 516, SE-75120 Uppsala, Sweden\\
% ...
$^{89}$ Zhengzhou University, Zhengzhou 450001, People's Republic of China\\		
		\vspace{0.2cm}
		$^{\dagger}$ Deceased\\
% ...
$^{c}$ Also at the Novosibirsk State University, Novosibirsk, 630090, Russia\\
% ...
}}

\begin{thebibliography}{**}

\bibitem{ref9}
G. Burdman, E. Golowich, J. L. Hewett, and S. Pakvasa,
\href{https://doi.org/10.1103/PhysRevD.52.6383}{Phys. Rev. D {\bf 52}, 6383 (1995)}.


\bibitem{ref11}
J.~M.~Dias, V.~R.~Debastiani, J.~J.~Xie, and E.~Oset,
\href{http://hepnp.ihep.ac.cn/article/doi/10.1088/1674-1137/42/4/043106}{Chin. Phys. C \textbf{42}, 4, 043106 (2018)}.

\bibitem{ref7}
B. Bajc, S. Fajfer, and R. J. Oakes,
\href{https://doi.org/10.1103/PhysRevD.51.2230}{Phys. Rev. D {\bf 51}, 2230 (1995)}.

\bibitem{ref8}
B. Bajc, S. Fajfer, and R. J. Oakes,
\href{https://doi.org/10.1103/PhysRevD.54.5883}{Phys. Rev. D {\bf 54}, 5883 (1996)}.





\bibitem{ref99}
S.~de Boer and G.~Hiller,
\href{https://link.springer.com/article/10.1007/JHEP08(2017)091}{JHEP \textbf{08}, 091 (2017)}.


\bibitem{ref13}
S. Fajfer and P. Singer,
\href{https://doi.org/10.1103/PhysRevD.56.4302}{Phys. Rev. D {\bf 56}, 4302 (1997)}.

\bibitem{ref14}
S. Fajfer, S. Prelovsek, and P. Singer,
\href{https://doi.org/10.1007/s100520050356}{Eur. Phys. J. C {\bf 6}, 471 (1999)}.

\bibitem{ref10}
H. Y. Cheng {\it et al}.,
\href{https://doi.org/10.1103/PhysRevD.51.1199}{Phys. Rev. D {\bf 51}, 1199 (1995)}.


\bibitem{ref88}
Y.~Cao and Q.~Zhao,
\href{https://journals.aps.org/prd/abstract/10.1103/PhysRevD.109.093005}{Phys. Rev. D  {\bf 109}, 093005 (2024)}.



\bibitem{ref12}
S. Fajfer, A. Prapotnik, S. Prelovsek, P. Singer, and J. Zupan,
\href{https://www.sciencedirect.com/science/article/pii/S0920563202019618?via\%3Dihub}{Nucl. Phys. Proc. Suppl. {\bf 115}, 93 (2003)}.



\bibitem{Biswas:2017eyn}
A.~Biswas, S.~Mandal, and N.~Sinha,
\href{https://www.worldscientific.com/doi/abs/10.1142/S0217751X18501944}{Int. J. Mod. Phys. A \textbf{33}, 1850194 (2018)}.

\bibitem{ref4}
B. Aubert {\it et al}. (BaBar Collaboration),
\href{http://dx.doi.org/10.1103/PhysRevD.78.071101}{Phys. Rev. D {\bf 78}, 071101(R) (2008)}.


\bibitem{ref6}
T. Nanut {\it et al}. (Belle Collaboration),
\href{http://dx.doi.org/10.1103/PhysRevLett.118.051801}{Phys. Rev. Lett. {\bf 118}, 051801 (2017)}.


\bibitem{ref3}
O. Tajima {\it et al}. (Belle Collaboration),
\href{http://dx.doi.org/10.1103/PhysRevLett.92.101803}{Phys. Rev. Lett. {\bf 92}, 101803 (2004)}.







\bibitem{cp}
R.~Aaij \textit{et al.} (LHCb Collaboration),
\href{https://journals.aps.org/prl/pdf/10.1103/PhysRevLett.122.211803}{Phys. Rev. Lett. {\bf 122}, 211803 (2019)}.

\bibitem{lhcb}
R. Aaij {\it et al}. (LHCb Collaboration),
\href{https://www.nature.com/articles/s41586-025-09119-3}{Nature {\bf 643}, 1223-1228 (2025)}.






\bibitem{lum_bes3}
M. Ablikim {\it et al.} (BESIII Collaboration),
\href{https://iopscience.iop.org/article/10.1088/1674-1137/37/12/123001}{Chin. Phys. C {\bf 37}, 123001 (2013);}
\href{https://doi.org/10.1016/j.physletb.2015.11.043}{Phys. Lett. B {\bf 753}, 629 (2016).}

\bibitem{lum_bes31}
M. Ablikim {\it et al.} (BESIII Collaboration),
\href{https://iopscience.iop.org/article/10.1088/1674-1137/ad70a0}{Chin. Phys. C {\bf 48}, 123001 (2024).}

\bibitem{BESIII}
M. Ablikim {\it et al.} (BESIII Collaboration),
\href{https://doi.org/10.1016/j.nima.2009.12.050}{Nucl. Instrum. Meth. A {\bf 614}, 345 (2010).}

\bibitem{Yu:IPAC2016-TUYA01}
C.~H.~Yu {\it et al.},
\href{https://doi.org/10.18429/JACoW-IPAC2016-TUYA01}{Proceedings of IPAC2016, Busan, Korea, 2016.}

\bibitem{cpc41}
M. Ablikim {\it et al.} (BESIII Collaboration),
\href{http://hepnp.ihep.ac.cn/en/article/doi/10.1088/1674-1137/44/4/040001}{Chin. Phys. C {\bf 44}, 040001 (2020).}


\bibitem{EcmsMea}
J.~Lu, Y.~Xiao, and X.~Ji,
\href{https://doi.org/10.1007/s41605-020-00188-8}{
Radiat. Detect. Technol. Methods {\bf 4}, 337–344 (2020).}

\bibitem{EventFilter}
J.~W.~Zhang, L.~H.~Wu, S.~S.~Sun {\it et al.},
\href{https://doi.org/10.1007/s41605-022-00331-7}
{Radiat. Detect. Technol. Methods {\bf 6}, 289 (2022).}


\bibitem{updatadata}
P. Cao {\it et al.} ,
\href{https://www.sciencedirect.com/science/article/abs/pii/S0168900219314068?via%3Dihub}{Nucl. Instrum. Meth. A 953, 163053 (2020).}


\bibitem{geant4}
S. Agostinelli {\it et al.} (GEANT4 Collaboration),
\href{https://doi.org/10.1016/S0168-9002(03)01368-8}{Nucl. Instrum. Meth. A {\bf 506}, 250 (2003).}

\bibitem{kkmc}
S. Jadach, B. F. L. Ward, and Z. Was,
\href{https://journals.aps.org/prd/pdf/10.1103/PhysRevD.63.113009} {Phys. Rev. D {\bf 63}, 113009 (2001);}
\href{https://doi.org/10.1016/S0010-4655(00)00048-5}{Comput. Phys. Commun. {\bf 130}, 260 (2000).}

\bibitem{evtgen1}
D.~J.~Lange,
\href{https://doi.org/10.1016/S0168-9002(01)00089-4} {Nucl. Instrum. Meth. A {\bf 462}, 152 (2001);}

\bibitem{evtgen2}
R.~G.~Ping,
\href{https://doi.org/10.1088/1674-1137/32/8/001}{Chin. Phys. C {\bf 32}, 599 (2008).}

\bibitem{pdg2022}
S. Navas {\it et al.} (Particle Data Group),
\href{https://journals.aps.org/prd/abstract/10.1103/PhysRevD.110.030001}{Phys. Rev. D {\bf 110}, 030001 (2024).}

\bibitem{lundcharm1}
J. C. Chen, G. S. Huang, X. R. Qi, D. H. Zhang, and Y. S. Zhu,
\href{https://journals.aps.org/prd/abstract/10.1103/PhysRevD.62.034003}{Phys. Rev. D {\bf 62}, 034003 (2000).}

\bibitem{lundcharm2}
R.~L.~Yang, R.~G.~Ping and H.~Chen,
\href{https://iopscience.iop.org/article/10.1088/0256-307X/31/6/061301}{Chin.\ Phys.\ Lett.\  {\bf 31}, 061301 (2014).}



\bibitem{photos}
  E.~Richter-Was,
  \href{https://doi.org/10.1016/0370-2693(93)90062-M}
 {Phys.\ Lett.\ B {\bf 303}, 163 (1993)}.





 \bibitem{mark3}
R. M. Baltrusaitis {\it et al.} (MARK III Collaboration),
\href{https://doi.org/10.1103/PhysRevLett.56.2140}{Phys. Rev. Lett. {\bf 56}, 2140 (1986);}
J. Adler {\it et al.} (MARK III Collaboration),
\href{https://doi.org/10.1103/PhysRevLett.60.89}{Phys. Rev. Lett. {\bf 60}, 89 (1988).}

\bibitem{ke2023}
B. C Ke {\it et al.},
\href{https://www.annualreviews.org/content/journals/10.1146/annurev-nucl-110222-044046}
{Ann. Rev. Nucl. Part. Sci. {\bf 73} 285-314 (2023)}.

\bibitem{cpc40}
M. Ablikim {\it et al.} (BESIII Collaboration),
\href{https://iopscience.iop.org/article/10.1088/1674-1137/40/11/113001}{Chin. Phys. C {\bf 40}, 113001 (2016).}



\bibitem{papernew1}
M.~Ablikim {\it et al.} (BESIII Collaboration),
\href{https://dx.doi.org/10.1103/PhysRevLett.124.241803}{Phys. Rev. Lett. {\bf 124}, 241803 (2020).}

\bibitem{kkpipi}
M.~Ablikim {\it et al.} (BESIII Collaboration),
\href{https://journals.aps.org/prd/abstract/10.1103/PhysRevD.102.052006}{Phys. Rev. D {\bf 102}, 052006 (2020).}

\bibitem{DCS-kpipi}
M.~Ablikim {\it et al.} (BESIII Collaboration),
\href{https://link.springer.com/article/10.1007/JHEP09(2022)107}{JHEP {\bf 09}, 107 (2022).}



\bibitem{ARGUS}
H. Albrecht {\it et al.} (ARGUS Collaboration),
\href{https://doi.org/10.1016/0370-2693(90)91293-K}{Phys. Lett. B {\bf 241}, 278 (1990).}


\bibitem{yuzhouxian}
M.~Ablikim {\it et al.} (BESIII Collaboration),
\href{https://www.sciencedirect.com/journal/physics-letters-b/vol/734}{Phys. Lett. B {\bf 277}, 734 (2014).}



\bibitem{AMP}
M. Ablikim {\it et al.} (BESIII Collaboration),
\href{https://journals.aps.org/prd/abstract/10.1103/PhysRevD.110.L111102}{Phys. Rev. D \textbf{110}, L111102 (2024).}




\bibitem{svs}
A.~Ryd {\it et al.},
\href{https://inspirehep.net/literature/707695}{EVTGEN-V00-11-07 (2005).}



\bibitem{bes3-pimuv}
M. Ablikim {\it et al.} (BESIII Collaboration),
\href{https://journals.aps.org/prl/abstract/10.1103/xj42-xgzf}{Phys. Rev. Lett. {\bf 135}, 091801 (2025).}




\bibitem{gammaselection}
M.~Ablikim {\it et al.} (BESIII Collaboration),
\href{https://docbes3.ihep.ac.cn/DocDB/0013/001303/024/JHEP11%282024%29119.pdf}{Phys. Rev. D {\bf 83}, 112005 (2011).}


\bibitem{bellebaipi}
W. Altmannshofer {\it et al.}(BelleII Collaboration),
\href{https://inspirehep.net/literature/1692393}{PTEP 2020, 029201 (2020).}

\bibitem{STCF1}
M. Achasov {\it et al.},
\href{https://link.springer.com/article/10.1007/s11467-023-1333-z}{Front. Phys. (Beijing) {\bf 19}, 14701 (2024).}

\bibitem{STCF2}
J. C. Bao {\it et al.},
\href{https://arxiv.org/pdf/2509.11522}{ arXiv: 2509.11522 [physics.acc-ph] (2025).}







\end{thebibliography}
\end{document}